\documentclass[aps,pre,twocolumn,groupedaddress,showpacs]{revtex4}
\usepackage{graphicx}
\usepackage{amssymb}

\begin{document}
\title{Effective temperatures and activated dynamics for a two-dimensional air-driven granular system on two approaches to jamming}

\author{A.R. Abate and D.J. Durian}
\affiliation{Department of Physics \& Astronomy, University of
Pennsylvania, Philadelphia, PA 19104-6396, USA}


\date{\today}

\begin{abstract}
We present experiments on several distinct effective temperatures in a granular system at a sequence of increasing packing densities and at a sequence of decreasing driving rates. This includes single-grain measurements based on the mechanical energies of both the grains and an embedded oscillator, as well as a collective measurement based on the Einstein relation between diffusivity and mobility, which all probe different time scales.  Remarkably, all effective temperatures agree.  Furthermore, mobility data along the two trajectories collapse when plotted vs effective temperature and exhibit an Arrhenius form with the same energy barrier as the microscopic relaxation time.
\end{abstract}



\maketitle


%
%

A critical challenge for the next decade is to understand nonequilibrium behavior such as commonalities in the jamming of glassy liquids, colloidal suspensions, and granular media~\cite{CMMP2010, LiuNagelBOOK}.  The concept of effective temperature, defined through the relation between the fluctuations and the response of a nonequilibrium system to small perturbation~\cite{Kurchan}, is an important unifying principle.  Away from thermal equilibrium, behavior could depend on history and driving, so one would not expect a unique effective temperature~\cite{ShokefLevine}.  Nevertheless, nearly ten different definitions yield a common low-frequency value in simulations of steadily driven Lennard-Jones systems~\cite{Barrat}, foams~\cite{LiuNagel}, and granular media~\cite{Makse}.  Furthermore, activated dynamics according to effective temperature have been incorporated into the theories of soft glassy rheology~\cite{SollichPRL97} and shear-transformation zones~\cite{Langer}, and have also been demonstrated in simulation~\cite{BarratEPL07, FalkPRL07, HaxtonPRL07}.

Despite these advances, the utility of effective temperatures in real-world systems is not as clear.  Violations of the fluctuation-dissipation relation have been reported for glass-forming materials~\cite{FirstTeff} and for granular systems~\cite{GranularTeff}, however no more than one effective temperature has ever been measured and compared for a single sample.  For aging colloidal glasses, there is even controversy about whether the Einstein relation between diffusivity and mobility gives an effective temperature any different from the bath temperature~\cite{AgingTeff}.  Furthermore, activated dynamics based on effective temperature have never been demonstrated in experiment.

In this paper we report on a quasi-two dimensional granular system of macroscopic spheres, which roll without slipping on a horizontal plane.  Uniform steady-state motion is excited by air, blown upward through a perforated mesh on which the balls roll.  Turbulent wakes are the source of both a rapidly-varying random driving force and of a velocity-dependent drag force on each sphere~\cite{RajeshNAT2004}; wake-wake interactions also cause a repulsive force between neighboring spheres~\cite{RajeshPRE2005ForcesonSpheres}.  This system exhibits such hallmark features as a growing plateau in mean-squared displacement~\cite{AbatePRE2006Multibead} and a growing dynamical correlation length~\cite{KeysNatPhys, AbatePRE07Topology} on approach to jamming.  Since the grains are macroscopic, and since the dynamics are slow, the ease of measurement and manipulation rivals that of computer simulations.  Here we take advantage of this feature, both for comparing different effective temperatures as well as for evaluating their meaning in terms of activated dynamics.

The experimental system consists of 50:50 bidisperse hollow polypropylene balls confined to a $12''$ diameter region; the diameters and masses are \{2.54~cm, 2.2~g\}  and \{2.86~cm, 3.0~g\}.  The balls are imaged from above at 30~Hz and 15~pixels/cm.  Position and velocity data are extracted using custom LabVIEW programs to respective accuracies of $0.001$~cm and $0.03$~cm/s.

\begin{figure}
\includegraphics[width=3.000in]{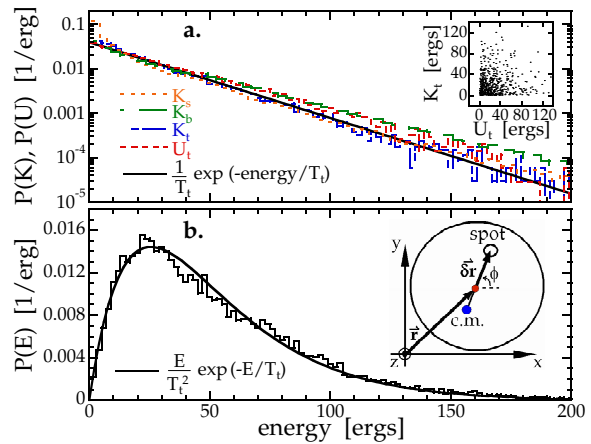}
\caption{(Color online) (a) Distribution of kinetic energies $K_s$ and $K_b$, for small and big balls, respectively, and distribution of kinetic energy $K_t$ and potential energy $U_t$ for the weighted-ball thermometer, at air speed $670$~cm/s and area density $0.43$. Inset: scatter plot of instantaneous kinetic and potential energy for the thermometer. (b) Distribution of thermometer total mechanical energy. Inset: the thermometer, as seen from above, is a weighted hollow sphere whose position degrees of freedom are measured by the coordinates {\bf r} of its geometric center and of a surface spot $\delta${\bf r} marked radially opposite the center of mass. The solid black curves in both (a,b) are the statistical mechanics predictions based on total average mechanical energy $T_t = \langle K_t + U_t \rangle /2$.}\label{KUE}
\end{figure}

The simplest effective temperature is based on kinetic energy, which has translational and rotational contributions since the balls roll without slipping.  Example results for the kinetic energy distributions of both size balls are plotted in Fig.~\ref{KUE}(a), for air speed $670$~cm/s and ball area density $0.43$.  The two distributions are similar and roughly exponential, as in a two-dimensional thermal system.  The average kinetic energy thus defines an effective ``granular'' temperature $T_g$.  At fixed airspeed, $T_g$ goes linearly to zero as the packing fraction increases toward 0.78, similar to Fig.~9 of Ref.~\cite{AbatePRE2006Multibead}.

The second effective temperature is inspired by theoretical consideration of a harmonically tethered test grain~\cite{Kurchan, MauriEPL06, BarratEPL07}.  This has been realized experimentally by optical trapping of a probe particle in an aging colloidal glass~\cite{GreinertPRL06}.  Here a similar effect is achieved with a 2.86~cm hollow plastic sphere that is partially filled with glue, so that the total mass is 3.8~g and the center-of-mass is 8.8~mm below the geometrical center.  When tilted, the weighted-sphere ``thermometer'' stores gravitational potential energy and experiences a restoring force like a pendulum.  The natural oscillation frequency is 2~Hz; the $Q$ factor is about ten; the r.m.s.\ tilt is less than $20^\circ$, so oscillations are harmonic.  A diagram of the thermometer is shown in Fig.~\ref{KUE}(b). The geometrical center of the thermometer and the location of the apex are tracked by video.  From these quantities we compute the total kinetic energy $K_t$, based on translational and rotational speeds plus mass and moment of inertia, as well as the total gravitational potential energy $U_t$, based on the rise in center-of-mass.

Example results for the thermometer energies are plotted in Fig.~\ref{KUE}.  The scatter plot of $K_t$ vs $U_t$ in the top inset reveals no correlations.  The probability distributions of $K_t$ and $U_t$ in the top plot are nearly indistinguishable, and appear to be exponential in reasonable agreement with the bath ball kinetic energy distributions under identical conditions.  This is characteristic of an object in thermal equilibrium with a heat bath, where the kinetic and potential degrees of freedom are independently populated and where the energies are exponentially distributed according to a Boltzmann factor.  This defines an effective temperature $T_t=\langle K_t+U_t\rangle/2$.  The thermal nature of the thermometer motion is further illustrated in the bottom plot, Fig.~\ref{KUE}(b), of the distribution of total thermometer energy $E=K_t+U_t$.  Since the system is two-dimensional, the density of states is proportional to $E$ and the distribution should be $P(E) = (E/{T_t}^2) \exp(-E/T_t)$, which matches the data very well.  The weighted sphere thus behaves like an object in thermal equilibrium, and $T_t$ is truly effective in its statistical mechanical meaning as a temperature.

\begin{figure}
\includegraphics[width=3.000in]{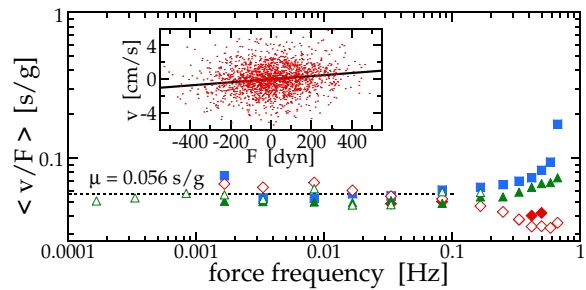}
\caption{(Color online) Mobility vs tilt frequency for $1''$ diameter test spheres at air speed $760$~cm/s and area density $0.57$.  Symbol types distinguish test spheres and tilt amplitudes; solid blue square: teflon, $18.45$~g, $0.52^{\circ}$;  hollow (solid) red diamond: acrylic, $10.12$~g, $2.35^{\circ}$ ($0.52^\circ$); hollow (solid) green triangle: wood, $6.4$~g, $0.52^{\circ}$ ($0.32^{\circ}$).  Inset: instantaneous velocity vs instantaneous force for the wood test sphere at $1.67\times 10^{-3}$~Hz and tilt amplitude $0.52^{\circ}$.  Mobility is the slope of the best fit to $v\propto F$ (black line).}\label{Mobs}
\end{figure}

The third effective temperature is based on the Einstein relation that diffusivity $D$ equals temperature times mobility $\mu$.  Since diffusion and flow involve grain-scale rearrangement of many neighboring beads, this probes long-time collective dynamics, as opposed to the short-time single-grain dynamics probed by $T_g$ and the intermediate-time dynamics probed by $T_t$.  Here we measure $D$ from the linear growth of the mean-squared displacement over at least one decade in time.  We measure $\mu$ for several 2.54~cm solid spheres with different masses, $m$, by tracking their motion while the entire bed is rocked sinusoidally in time.  When the system is at angle $\theta$ away from horizontal, a test sphere experiences a force $F$ down the plane given by $mg\sin\theta$, minus the buoyancy due to density difference with bath balls, plus a contribution set by the in-plane acceleration due to the location of rotation axis at the bottom of the windbox about four feet below the balls.  Therefore, parallel to the tilt direction, a test sphere acquires an average drift speed of $v=F\mu$ that is superposed on its random thermal motion.  To extract mobility, we therefore make a scatter plot of parallel instantaneous speed vs instantaneous force at all times and then we fit for the proportionality constant, as in the inset of Fig.~\ref{Mobs}.  Mobility values for several test spheres are plotted vs tilt frequency in the main plot.  If the stochastic motion of the bath balls is unperturbed, the mobility results are independent of test mass, tilt frequency, and tilt amplitude; this demonstrates linearity of response.  Thus we reliably measure both mobility and diffusivity, from which we compute an effective ``Einstein'' temperature $T_e = D/\mu$.

\begin{figure}
\includegraphics[width=3.000in]{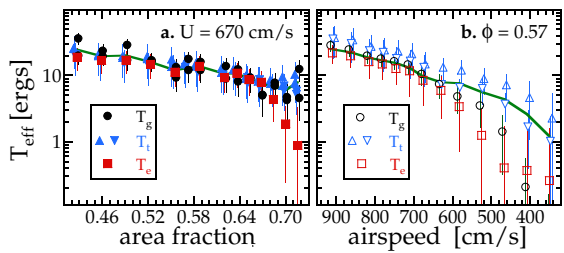}
\caption{(Color online) Effective temperatures vs (a) area fraction at airspeed 670~cm/s, and (b) vs driving airspeed at area fraction 0.57.  The granular temperature $T_g$ is based on the ball kinetic energies; the thermometer temperature $T_t$ is based on the mechanical energy of weighted-sphere oscillators with two different masses, 3.83~g for upward triangles and 6.82~g for downward triangles; the Einstein temperature $T_e$ is the ratio of diffusivity to mobility.  The solid green curve is the weighted average of all these measures.  Measurement uncertainties for $T_g$ and $T_t$ are comparable to symbol size, while the uncertainties in $T_e$ are indicated by error bars.}\label{Tjam}
\end{figure}

Results for the three effective temperatures are compared in Fig.~\ref{Tjam} as the system is brought closer to jamming along two different trajectories.  For increasing area fraction at fixed airspeed $670$~cm/s, and also for decreasing airspeed at a fixed area fraction $0.57$, the various effective temperatures are found to decrease by roughly one order of magnitude.  However, the more crucial feature of Fig.~\ref{Tjam} is that the values of all three types of effective temperature are in agreement to within uncertainties estimated from the imaging resolution and fitting accuracies.  Close to jamming the Einstein temperature systematically drops below the others, but this may be because the growing sub-diffusive plateau causes an underestimate of $D$.  But overall, the good agreement allows us to average together $\{T_g, T_t, T_e\}$ with weights given by the uncertainties.  The result is a single effective temperature, $T_{\rm eff}$, displayed by solid green curves in Fig.~\ref{Tjam}.

\begin{figure}
\includegraphics[width=3.000in]{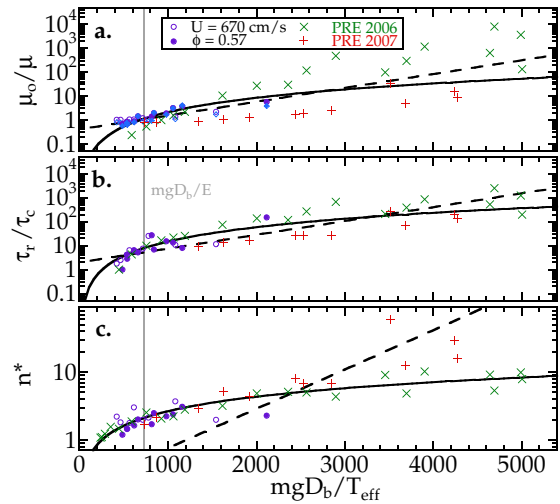}
\caption{(Color online) (a) Inverse mobility scaled by $\mu_o=\tau_c/m$, (b) dimensionless extent of subdiffusive plateau in mean-squared displacement, and (c) average number of grains in kinetic heterogeneities, all vs inverse effective temperature scaled by ball weight times diameter.  Open and closed symbols are for the polypropylene balls used in Figs.~1-3, along two trajectories as labeled.  The $\times$ symbols are based on earlier data \protect{\cite{AbatePRE2006Multibead}} for a bidisperse mixture of 0.635 and 0.873~cm diameter steel balls, and the $+$ symbols are based on earlier data \protect{\cite{AbatePRE07Topology}} for a bidisperse mixture of 0.318 and 0.397~cm diameter steel balls; for both, jamming is approached by change in packing fraction at constant airspeed.  In (a) the diamond, $\times$, and $+$ symbols represent $T_{\rm eff}/D$.  In (a-c) the dashed lines are Arrhenius fits, $\propto \exp(E/T_{\rm eff})$, all with the same energy barrier indicated by the vertical gray line.  The solid curves are power-law fits $\mu_o/\mu \propto \tau_r/\tau_c \propto (mgD_b/T_{\rm eff})^{2.0\pm0.5}$ and $n^* \propto (mgD_b/T_{\rm eff})^{0.7\pm0.2}$}\label{Arrhenius}
\end{figure}

We now examine the mobility, not as a function of the actual experimental control parameters of area fraction and airspeed, but rather as a function of the effective temperature.  As usual for glass-forming liquids, we make a semi-logarithmic plot of $1/\mu$ vs $1/T_{\rm eff}$ in Fig.~\ref{Arrhenius}(a).  The values of $T_{\rm eff}/D$, also included, differ slightly from $1/\mu$ because the Einstein temperature is not identical to $T_{\rm eff}$.  Here the temperature axis is scaled by ball weight times diameter, $mgD_b$; the mobility axis is scaled by $\mu_o=\tau_c/m$, where $\tau_c$ is the cage-recognition time when grains cross from ballistic to subdiffusive motion, as defined in Fig.~11 of Ref.~\cite{AbatePRE2006Multibead}.  Unexpectedly, we find that the data along the two different trajectories collapse when plotted according to effective temperature.  Scaled $T_g/D$ results for larger collections of steel balls \cite{AbatePRE2006Multibead, AbatePRE07Topology} also collapse nicely, though with more scatter.  The ``thermodynamic'' state of air-driven grains, and their distance from jamming, are entirely captured by the value of $T_{\rm eff}/(mgD_b)$ without regard to the actual control parameters of airspeed and density.  Next, remarkably, we find that $1/\mu$ is exponential in $E/T_{\rm eff}$ and hence is Arrhenius, just like in strong glass-forming liquids where relaxation is activated with a single energy barrier $E$.  Here the value of $mgD_b/E$, indicated by the vertical gray line in Fig.~\ref{Arrhenius}, is set by air-mediated repulsion of a grain by its surrounding cage of neighbors.  Note $\mu\approx\mu_o$ when $T_{\rm eff}\approx E$, supporting the Arrhenius interpretation.

While mobility can be thought of as a relaxation rate, it is instructive to compare with more direct microscopic time and length scales.  In particular, the proximity to jamming may also be gauged from the mean-square displacement by the separation of the cage-recognition time $\tau_c$ and the rearrangement time $\tau_r$, when grains cross from subdiffusive to diffusive motion.  The rearrangement dynamics are spatially heterogeneous, punctuated by sudden motion of a string-like cluster of neighboring grains \cite{SHD} whose average number $n^*$ may be deduced from four-point correlation functions \cite{AbatePRE07Topology}.  Previously we measured both $\tau_r/\tau_c$ \cite{AbatePRE2006Multibead, KeysNatPhys, AbatePRE07Topology} and $n^*$ \cite{AbatePRE07Topology} vs packing fraction at fixed airspeed for larger collections of smaller steel balls.  Now we compute these quantities for all systems, and plot the results vs $mgD_b/T_{\rm eff}$ in Figs.~\ref{Arrhenius}(c,d).  As $mgD_b/T_{\rm eff}$ increases on approach to jamming, the extent $\tau_r/\tau_c$ of sub-diffusive motion and the size $n^*$ of heterogeneities both collapse for the different grain sizes and also appear to grow without bound.  Arrhenius fits with the same energy barrier as for mobility are satisfactory for $\tau_r/\tau_c$, but not for $n^*$, both as expected.  Power-law fits are also satisfactory, as shown.  The latter give $\tau_r/\tau_c \sim (mgD_b/T_{\rm eff})^{2.0\pm0.5}$ and $n^*\sim (mgD_b/T_{\rm eff})^{0.7\pm0.2}$; these exponents are consistent with simulation of a Lennard-Jones liquid~\cite{DonatiPRL99}.  Intriguingly, the size exponent is also consistent with simulations of athermal systems vs packing density~\cite{tzero}.

In summary, we have presented the first experimental evidence that distinct effective temperatures can agree.  This holds across a range of conditions, and includes measures based on the kinetic energies of two sized bath spheres at short times, the kinetic and potential energies of two weighted-ball oscillators at intermediate times, and the ratio of diffusivity to mobility at long times.  For a nonequilibrium system there is no {\it a priori} reason for these seven quantities to agree.  Furthermore we have demonstrated that $T_{\rm eff}$ acts as the sole state variable for air-driven grains and that relaxation times are Arrhenius in $T_{\rm eff}$.  Even more than the measures of structure and dynamics in \cite{AbatePRE2006Multibead}, and the measures of spatially-heterogeneous dynamics in \cite{KeysNatPhys, AbatePRE07Topology}, this establishes air-fluidized beads as a faithful model for the glass transition.  Besides extending the universality of the jamming and effective temperature concepts, this is also important because key microscopic quantities such as $\tau_r/\tau_c$ and $n^*$ are not as fully accessible in other experimental systems.  One interesting line of future research would be to impose shear, or an unlikely initial configuration, so that the system is away from ``equilibrium'', and to study differences in the various effective temperatures.  An even broader line would be to explore when and why driven systems exhibit equilibrium-like behavior~\cite{driventhermo}.

We thank J.-L.\ Barrat, A.J.\ Liu, S.\ Teitel, and Y.\ Shokef for helpful discussions. Our work was supported by the NSF through grant DMR-0704147.


\end{document}